\documentclass{webofc}
\usepackage[varg]{txfonts}   % Web of Conferences font
\begin{document}
\title{Integrating the PanDA Workload Management System with the Vera C. Rubin Observatory}
%
% subtitle is optionnal
%
%%%\subtitle{Do you have a subtitle?\\ If so, write it here}

\author{\firstname{Edward} \lastname{Karavakis}\inst{1}\fnsep\thanks{\email{Edward.Karavakis@cern.ch}} \and
        \firstname{Wen} \lastname{Guan}\inst{1}\and
        \firstname{Zhaoyu} \lastname{Yang}\inst{1}\and
        \firstname{Tadashi} \lastname{Maeno}\inst{1}\and
        \firstname{Torre} \lastname{Wenaus}\inst{1} \and
        \firstname{Jennifer } \lastname{Adelman-McCarthy}\inst{2}\and
        \firstname{Fernando} \lastname{Barreiro Megino}\inst{3}\and
        \firstname{Kaushik} \lastname{De}\inst{3} \and
        \firstname{Richard} \lastname{Dubois}\inst{4} \and
        \firstname{Michelle} \lastname{Gower}\inst{5} \and
        \firstname{Tim} \lastname{Jenness}\inst{6} \and
        \firstname{Alexei} \lastname{Klimentov}\inst{1} \and
        \firstname{Tatiana} \lastname{Korchuganova}\inst{7} \and
        \firstname{Mikolaj} \lastname{Kowalik}\inst{5} \and
        \firstname{Fa-Hui} \lastname{Lin}\inst{3}\and
        \firstname{Paul} \lastname{Nilsson}\inst{1}\and
        \firstname{Sergey} \lastname{Padolski}\inst{1}\and
        \firstname{Wei} \lastname{Yang}\inst{4}\and
        \firstname{Shuwei} \lastname{Ye}\inst{1}
}

\institute{Brookhaven National Laboratory, Upton, NY, USA
\and
           Fermi National Accelerator Laboratory, Batavia, IL, USA
\and
           University of Texas at Arlington, Arlington, TX, USA 
\and
           SLAC National Accelerator Laboratory, Menlo Park, CA, USA
\and
           National Center for Supercomputing Applications, Urbana, IL, USA
\and
           Vera C.\ Rubin Observatory, Tucson, AZ, USA
\and
           University of Pittsburgh, Pittsburgh, PA, USA
          }

\abstract{%
The Vera C. Rubin Observatory will produce an unprecedented astronomical data set for studies of the deep and dynamic universe. Its Legacy Survey of Space and Time (LSST) will image the entire southern sky every three to four days and produce tens of petabytes of raw image data and associated calibration data over the course of the experiment's run. More than 20 terabytes of data must be stored every night, and annual campaigns to reprocess the entire dataset since the beginning of the survey will be conducted over ten years. The Production and Distributed Analysis (PanDA) system was evaluated by the Rubin Observatory Data Management team and selected to serve the Observatory’s needs due to its demonstrated scalability and flexibility over the years, for its Directed Acyclic Graph (DAG) support, its support for multi-site processing, and its highly scalable complex workflows via the intelligent Data Delivery Service (iDDS). PanDA is also being evaluated for prompt processing where data must be processed within 60 seconds after image capture. This paper will briefly describe the Rubin Data Management system and its Data Facilities (DFs). Finally, it will describe in depth the work performed in order to integrate the PanDA system with the Rubin Observatory to be able to run the Rubin Science Pipelines using PanDA.
}
\maketitle
\section{Introduction}
\label{intro}
The Vera C. Rubin Observatory's Legacy Survey of Space and Time (LSST) \cite{rubin} will conduct a ten-year survey of the Southern Hemisphere sky, producing an unparalleled dataset for studying the universe. LSST's imaging capabilities and nightly data collection will generate a significant volume of raw image data and associated calibration data, creating data processing and management challenges.

This paper provides an overview of the Rubin Observatory, including the LSST and its data generation and processing challenges, and it introduces the PanDA Workload Management System \cite{panda}, highlighting its scalability and flexibility. Finally, it focuses on the integration of PanDA with the Rubin Observatory \cite{DMTN-168}, including the mapping of a Rubin Directed Acyclic Graph (DAG) to PanDA workload, multi-site processing, PanDA deployment at SLAC Kubernetes, monitoring, near real-time log access, and prompt processing considerations. 

\section{Vera C. Rubin Observatory}
\label{rubin}
\subsection{The Legacy Survey of Space and Time (LSST)}
\label{rubin-lsst}
Rubin Observatory's LSST is a project that aims to survey the entire Southern Hemisphere sky every three to four days, capturing a substantial volume of data. Equipped with a 3.2~gigapixel camera \cite{2010SPIE.7735E..0JK}, the Observatory's telescope will image the entire visible sky, allowing for the detection of objects that have changed in brightness, such as supernovae, or in position, such as asteroids. The Observatory's light-collecting power and sensitive camera are expected to discover approximately 20 billion galaxies and an equal number of stars.

The data collection of LSST is substantial, with each night producing over 20 terabytes of raw image data and associated calibration data. Over the course of its ten-year mission, LSST is estimated to generate approximately 60 petabytes of raw image data and several hundreds of petabytes of science-ready images and a multi-petabyte relational database to host the astronomical catalog. The scale and complexity of this dataset present significant challenges in terms of data processing, storage, and management.

\subsection{Data Butler}
\label{rubin-butler}
The Data Butler \cite{butler} serves as the interface between the data and the pipeline tasks, facilitating seamless access to the substantial volume of data generated by the Observatory. Data Butler is responsible for data management, providing functionalities for data ingestion, organization, metadata management, and retrieval.

At its core, Data Butler acts as a data repository, storing and organizing the datasets acquired by the Observatory. It ensures the integrity and traceability of the data, allowing for efficient and reliable data processing workflows. Data Butler's metadata management capabilities enable efficient querying and filtering of data based on various criteria, such as time, location, or specific scientific requirements.

In addition to data management, there is an execution framework that uses the facilities of the Data Butler to simplify the interface between algorithmic code and job execution. It provides the necessary inputs and parameters to the tasks, allowing them to operate on the relevant data subsets. Data Butler's ability to handle complex dependencies and manage the flow of data between tasks ensures the efficient execution of the data processing workflows.

Furthermore, Data Butler enables data provenance and reproducibility by keeping track of the processing history and maintaining a record of all modifications and operations performed on the data. This provenance tracking is vital for scientific reproducibility and quality control, allowing researchers to trace back to the original data sources and understand the steps involved in generating the final scientific products.

\subsection{Batch Production Service (BPS)}
\label{rubin-bps}
BPS \cite{bps} is the component of the Rubin Observatory pipeline execution system responsible for providing an interface to several Workload Management Systems (WMSs). It
acts as a bridge between the data management capabilities of Data Butler and the execution capabilities of various WMSs, such as PanDA, HTCondor, Parsl, and Pegasus, utilizing plugins to facilitate the seamless integration and coordination between these components. It orchestrates and manages the batch processing of data, ensuring seamless coordination between Data Butler and the WMS. It facilitates the retrieval of required data subsets from Data Butler, delivers them to the WMS for processing, and handles essential tasks such as data movement, job generation, and task scheduling. 

\subsection{Data Facilities}
\label{rubin-dfs}
To address the data processing and management challenges of LSST, the Observatory has established multiple data facilities \cite{lsstoverview}. These facilities include the United States Data Facility (USDF), the French Data Facility (FrDF), the UK Data Facility (UKDF), and the Interim Data Facility (IDF) hosted on the Google Cloud Platform (GCP).

USDF, located at the SLAC Shared Scientific Data Facility (S3DF) \cite{s3df} at the SLAC National Accelerator Laboratory, plays a crucial role in LSST's data processing and storage. It handles all prompt processing activities, 35\% of the data release processing, and provides data access services to the US and international research community. FrDF, hosted at the CC-IN2P3 in Lyon, France, contributes to 40\% of the data release processing and serves as a backup for raw data and select published products. UKDF, hosted by IRIS and GridPP in the UK, handles 25\% of the data release processing.

IDF serves as a cloud-based interim facility used for pre-operations activities. It provides computational resources and storage for early data processing, algorithm development, and testing. IDF provided a platform to integrate the Rubin and PanDA software systems before the compute hardware was installed at SLAC.
It was used to demonstrate the viability of the system by processing all the data for the Data Preview 0 exercise \cite{RTN-041}.

\section{Production and Distributed Analysis (PanDA) Workload Management System}
\label{panda}
PanDA is a highly scalable and flexible workload management system \cite{pandadistibuted} that was initially developed to address the computational challenges of the ATLAS experiment \cite{atlas} at CERN. PanDA has demonstrated its capability to manage 24x365 processing on approximately 800,000 concurrent cores globally for ATLAS, encompassing all types of workflows, resource types, and a large user base. Its horizontal scaling capabilities have been further enhanced with support for Kubernetes (K8s) \cite{k8s} deployment.

% \begin{figure}[h]
% % Use the relevant command for your figure-insertion program
% % to insert the figure file.
% \centering
% \includegraphics[width=12cm]{images/ATLASPanDA.png}
% \caption{Slots of PanDA running jobs for ATLAS during 2022. }
% \label{fig_atlas}       % Give a unique label
% \end{figure}

PanDA offers an easy-to-use submit client over the command line interface or as a Python API, allowing users to submit their scientific workflows through a Python script or a Jupyter notebook. With PanDA's ability to smoothly scale horizontally, it is well-suited to handle the anticipated demands of the Rubin Observatory, with about 100,000 concurrent jobs in 3 geographically distributed data facilities.

Scalability is a crucial aspect of PanDA's design, allowing it to effectively manage large-scale scientific workloads. The system seamlessly integrates with various computing resources, including clusters, grids, and clouds, and can dynamically allocate resources based on workload demands. This flexibility ensures efficient resource utilization, minimizing idle time and maximizing productivity.

Furthermore, PanDA's scalability extends beyond its ability to handle increasing computational demands. It also supports the growth of user communities, accommodating diverse research groups and institutions. With over 1,500 users and approximately 300 million jobs per year within the ATLAS collaboration, PanDA has proven its ability to support large-scale scientific collaborations effectively.

\section{Integration of PanDA with the Vera C.\ Rubin Observatory}
\label{integration}
\begin{figure}[h]
% Use the relevant command for your figure-insertion program
% to insert the figure file.
\centering
\includegraphics[width=12cm]{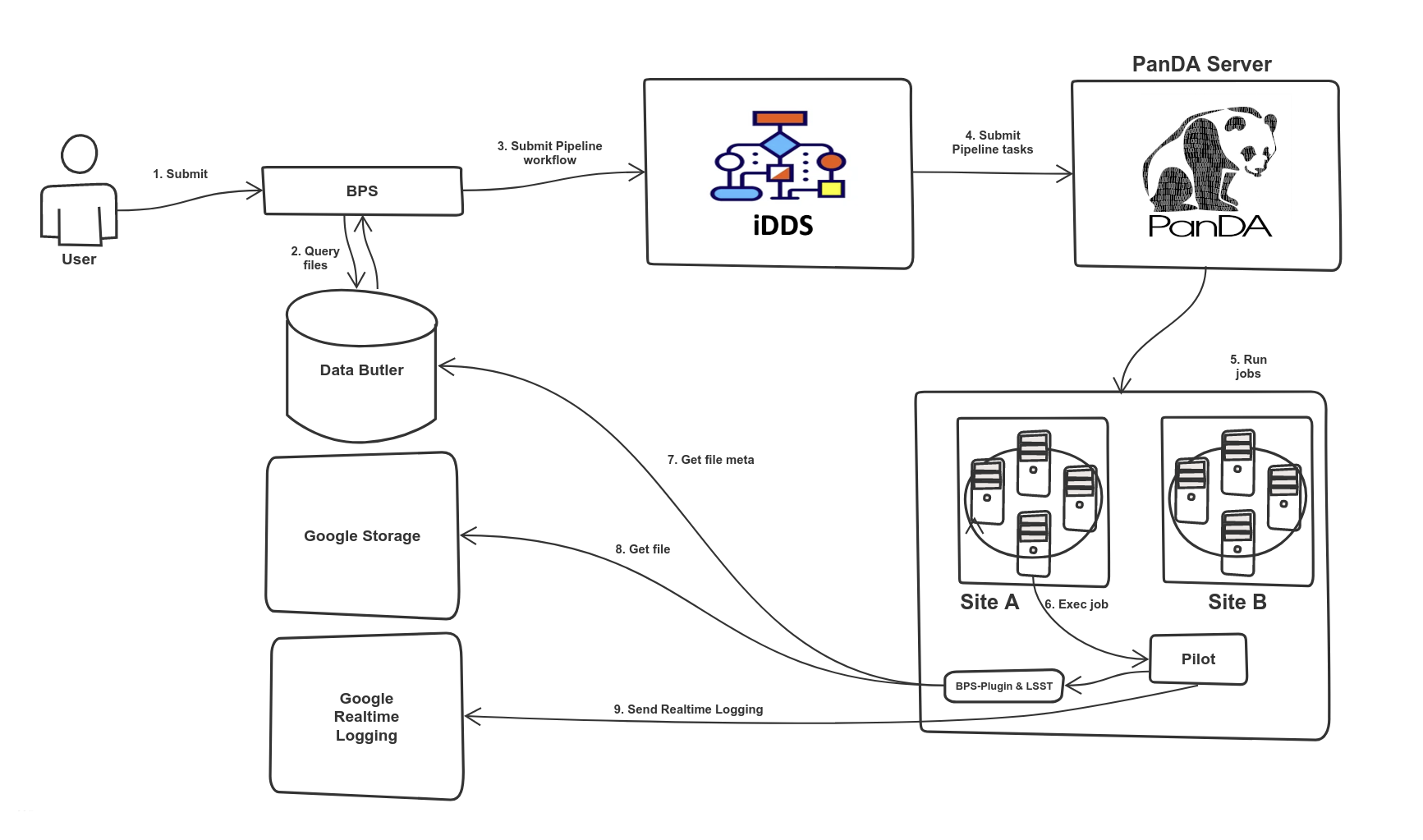}
\caption{PanDA integration with Rubin Observatory.}
\label{fig_integration}       % Give a unique label
\end{figure}

Figure \ref{fig_integration} illustrates the integration of PanDA within the data processing workflow of the Rubin Observatory.\footnote{A user guide is available at \url{https://panda.lsst.io}.} The core components of this integration include the LSST Science Pipelines \cite{2019ASPC..523..521B}, which consist of the algorithmic payloads and the Data Butler and pipeline framework. Data Butler serves as the interface between the data and the pipeline tasks, providing data access and facilitating seamless interaction between the two.

BPS acts as an intermediary between Rubin's pipeline execution system and PanDA, serving as the interface for integrating Rubin with PanDA and the iDDS client. PanDA, as the workload management system, manages and schedules the Rubin workload across distributed resources. The PanDA pilot, integrated with the Rubin Data Butler access, facilitates the execution of the Rubin workload while also providing access to Google storage and enabling real-time logging. The storage of pilot logs in GCP, as required by Rubin, enables efficient storage and real-time monitoring of processing tasks. It's worth noting that PanDA can also route logs to other logging platforms if necessary.

Community engagement and collaboration was fostered through regular weekly meetings between all relevant teams, ensuring effective coordination for essential development work.

\subsection{Mapping Rubin DAG to PanDA Workload}
\label{integration-dag}
The Rubin Science Pipelines, which form the core data processing framework for the Rubin Observatory, are composed of various pipeline tasks organized in a DAG structure. These DAGs describe the dependencies between tasks and the flow of data processing. In the context of the Rubin Observatory, LSST's processing workflows are modeled as quantum graphs, where each node represents the execution of a science algorithm with a well known set of inputs and outputs. The integration of PanDA with the Observatory involves mapping these DAGs to PanDA workflows, enabling efficient and optimized execution of the Observatory's processing tasks, taking advantage of PanDA's scalability and workload management capabilities.

The intelligent Data Delivery Service (iDDS) \cite{idds} plays a crucial role in supporting complex, fine-grained workflows defined through DAGs or workflow description languages. iDDS has been successfully utilized in the ATLAS experiment for the data carousel activity \cite{datacarousel} and various machine learning and analysis workflows \cite{machinelearning}.

\subsection{Multi-Site Processing}
\label{integration-siteproc}
Rubin Observatory's data processing requirements span multiple data facilities. PanDA's support for multi-site processing is particularly valuable in this context, allowing efficient coordination and workload distribution across different sites.

To enable multi-site processing for Rubin \cite{DMTN-213}, both the Rubin Data Management middleware and PanDA/iDDS required development in order to ensure that the quantum graph and execution Data Butler, created at one data facility, can be seamlessly used at another facility. Additionally, mechanisms for merging outputs and metadata back to the main Data Butler registry  with the help of Rucio \cite{rucio} were developed to support efficient processing across multiple sites.

\subsection{PanDA Deployment on Kubernetes}
\label{integration-k8s}
To support the integration of PanDA with the Rubin Observatory, a deployment of PanDA at the SLAC National Accelerator Laboratory was carried out, utilizing Kubernetes (K8s) as the underlying container orchestration platform. The deployment included all the necessary components of PanDA, such as the PanDA Server, JEDI, Indigo IAM authentication, Harvester, iDDS, PanDA Monitor, and ActiveMQ.

In addition, a highly available PostgreSQL database cluster, based on the CloudNativePostgreSQL (CNPG) \cite{cnpg} framework, was deployed to ensure the reliability and scalability of the database backend for PanDA. The use of Kubernetes and CNPG provides essential cloud-native capabilities, such as self-healing, high availability, rolling updates, and resource management, ensuring the stability and performance of the PanDA deployment at SLAC.

\subsection{Monitoring and Near Real-Time Log Access}
\label{integration-monitoring}
Effective monitoring is crucial for ensuring the smooth operation of the Rubin data processing workflows through PanDA. To enhance the monitoring capabilities, PanDA Monitor \cite{pandamonitor} was extended to support Rubin job monitoring.

\begin{figure}[h]
% Use the relevant command for your figure-insertion program
% to insert the figure file.
\centering
\includegraphics[width=12cm]{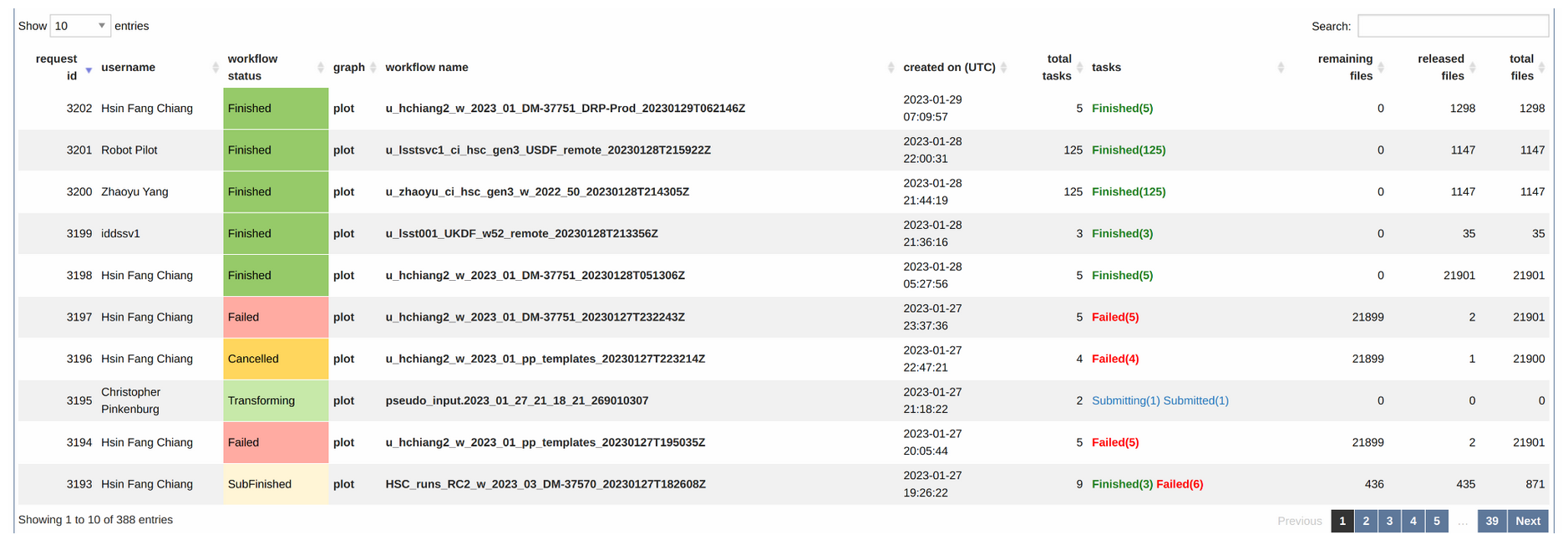}
\caption{Workflow monitoring in PanDA Monitor allowing hierarchical navigation. }
\label{fig_pandamon}       % Give a unique label
\end{figure}

As seen in Figure \ref{fig_pandamon}, PanDA Monitor provides hierarchical navigation at different levels, allowing users to track the progress of workflows, tasks, jobs, and logs. Additionally, memory usage monitoring using the prmon \cite{prmon} tool, originally developed by ATLAS, was integrated into PanDA Monitor. This enhancement provides valuable insights into memory consumption during job execution.

Furthermore, the integration of near real-time log access into the PanDA environment offers a significant advantage for monitoring and debugging purposes. The Pilot captures payload logs and sends them as JSON to the Google Cloud Logging service \cite{googlecloudlog}, allowing direct access to logs from PanDA Monitor. This near real-time log access complements conventional log access, providing additional information for troubleshooting and performance analysis.

\subsection{PanDA for Prompt Processing}
\label{integration-prompt}
Prompt processing is a critical aspect of the Rubin Observatory's operations, involving the rapid processing of data and issuing alerts within a short timeframe of the order of a minute. PanDA is currently being evaluated for prompt processing within the Observatory, with the aim of enabling the initiation of processing within seconds on dedicated resources at SLAC. 

To achieve prompt processing, several developments were implemented. These include the use of semi-persistent pilots running on worker nodes, task resurrection through notifications to skip unnecessary overhead, direct job communication via ActiveMQ, and a communication channel between the JEDI component and the PanDA server. These advancements significantly reduce latency and support near-real-time data processing, making prompt processing feasible for the Rubin Observatory.

\section{Results}
\label{results}
\begin{figure}[h]
% Use the relevant command for your figure-insertion program
% to insert the figure file.
\centering
\includegraphics[width=12cm]{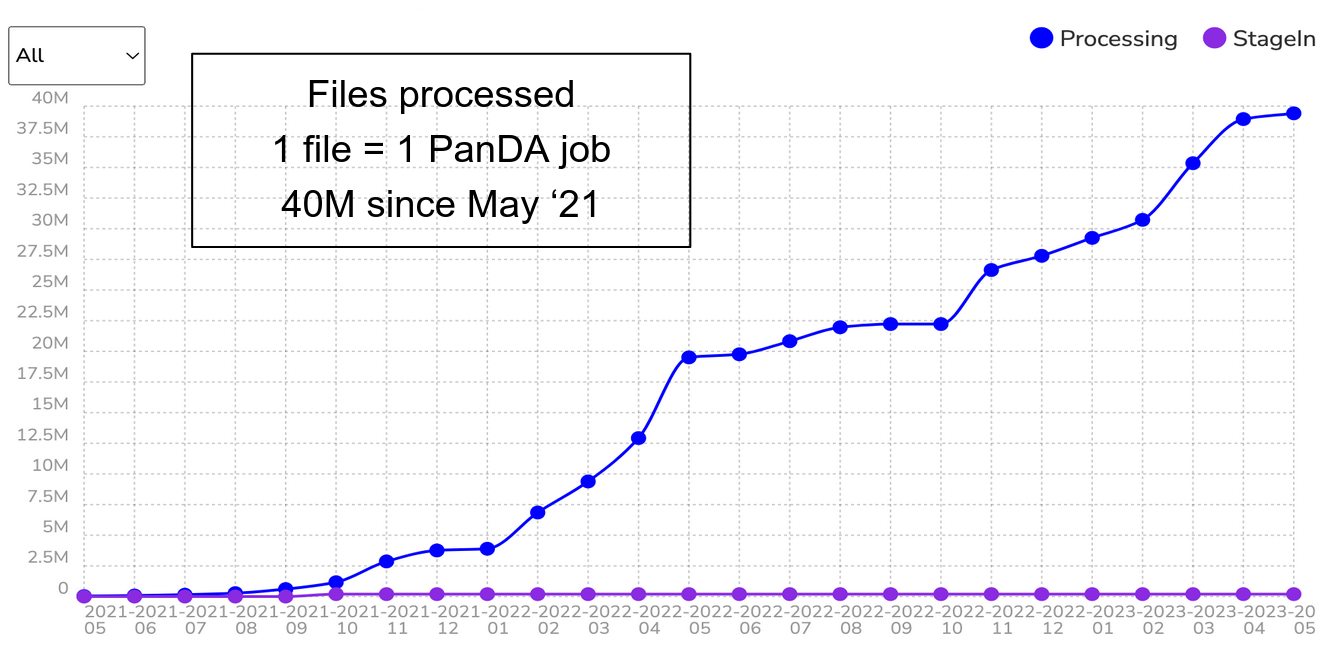}
\caption{Accumulation of the number of Rubin Observatory jobs managed by PanDA since May 2021. }
\label{fig_rubin}       % Give a unique label
\end{figure}

During Phase 2 of the Rubin Observatory's Data Preview 0 (DP0.2) in 2021 \cite{RTN-041}, PanDA demonstrated the capability to run 16 million jobs at the Google-based IDF \cite{RTN-039}. Most jobs were processed on a cluster with approximately 4000 cores, up to 14GB/core RAM with a total CPU usage of 2.5M core-hours. Eight million jobs were also processed for the Hyper-SuprimeCam (HSC) reprocessing at USDF in SLAC. 

The successful processing of DP0.2 drove the decision to endorse PanDA for the Data Release Production (DRP) campaigns \cite{RTN-013} and since 2022, Rubin Observatory has been using PanDA for the DRP campaigns. Figure \ref{fig_rubin} shows the number of Rubin jobs processed with PanDA since May 2021. For the 2023 DRP campaigns, it is estimated to have around 36 million jobs for the HSC Public Data Release 2 (HSC-PDR2) and around 8M for the HSC reprocessing.

\section{Conclusion}
\label{conclusion}
The integration of the PanDA WMS with the Rubin Observatory has proven to be successful in enabling efficient and scalable data processing for LSST. Through successful deployments and campaigns, such as DP0.2 and HSC reprocessing, PanDA has demonstrated its reliability and performance in meeting the processing requirements of the Rubin Observatory. The integration efforts have paved the way for seamless execution of Rubin Science Pipelines, efficient multi-site processing, improved monitoring and log access, and prompt processing capabilities in order to be able to handle the significant amount of data generated by the Observatory.

The integration of PanDA with the Rubin Observatory serves as a testament to the value of collaborative efforts between scientific facilities and distributed computing projects. It showcases the successful integration of advanced workload management systems with complex astronomical data processing workflows. Most importantly, the PanDA development work undertaken for Rubin Observatory is experiment agnostic and is currently used by other experiments as well, benefiting the wider scientific community.

\section{Acknowledgments}
\label{acknowledgments}

This manuscript has been authored by employees of Brookhaven Science Associates, LLC under Contract No.\ DE-SC0012704 with the U.S. Department of Energy. The publisher by accepting the manuscript for publication acknowledges that the United States Government retains a nonexclusive, paid-up, irrevocable, worldwide license to publish or reproduce the published form of this manuscript, or allow others to do so, for United States Government purposes.

This material is based upon work supported in part by the National Science Foundation through Cooperative Agreement AST-1258333 and Cooperative Support Agreement AST-1202910 managed by the Association of Universities for Research in Astronomy (AURA), and the Department of Energy under Contract No.\ DE-AC02-76SF00515 with the SLAC National Accelerator Laboratory managed by Stanford University. Additional Rubin Observatory funding comes from private donations, grants to universities, and in-kind support from LSSTC Institutional Members.

% BibTeX or Biber users please use (the style is already called in the class, ensure that the "woc.bst" style is in your local directory)
\bibliography{bib}
%
% Non-BibTeX users please use
%

\end{document}